\documentclass[aps,pra,groupedaddress,showpacs,twocolumn]{revtex4}%
\usepackage{amsfonts}
\usepackage{amsmath}
\usepackage{amssymb}
\usepackage{graphicx}%
\providecommand{\U}[1]{\protect\rule{.1in}{.1in}}

\begin{document}
\preprint{HEP/123-qed}
\title{Level shift and decay dynamics of a quantum emitter around plasmonic nanostructure}
\author{Meng Tian}
\affiliation{College of Physical Science and Mechanical Engineering, Jishou University, Jishou 416000, China }
\author{Yong-Gang Huang}
\email{huang122012@163.com}
\affiliation{College of Physical Science and Mechanical Engineering, Jishou University, Jishou 416000, China }
\author{Sha-Sha Wen}
\affiliation{College of Physical Science and Mechanical Engineering, Jishou University, Jishou 416000, China }
\author{Xiao-Yun Wang}
\affiliation{College of Physical Science and Mechanical Engineering, Jishou University, Jishou 416000, China }
\author{Hong Yang}
\affiliation{College of Physical Science and Mechanical Engineering, Jishou University, Jishou 416000, China }
\author{Jin-Zhang Peng}
\affiliation{College of Physical Science and Mechanical Engineering, Jishou University, Jishou 416000, China }
\author{He-Ping Zhao}
\affiliation{College of Physical Science and Mechanical Engineering, Jishou University, Jishou 416000, China }

\pacs{22}

\begin{abstract}
We put forward a general approach for calculating the quantum energy level shift for emitter in
arbitrary nanostructures, in which the energy level shift is expressed by the sum of the real part of
the scattering photon Green function (GF) and a simple integral about the imaginary part of the photon GF in the
real frequency range without principle value. Compared with the method of direct principal value integral over the positive frequency axis and the method by transferring into the imaginary axis, this method avoids the principle value integral and the calculation of the scattering GF with imaginary frequency. In addition, a much narrower frequency range about the scattering photon GF in enough to get a convergent result. It is numerically demonstrated in the case for a quantum emitter (QE) located around a nanosphere and in a gap plasmonic nanocavity. Quantum dynamics of the emitter is calculated by the time domain method through solving Schr\"{o}dinger equation in the form of Volterra integral of the second kind and by the frequency domain method based on the Green's function expression for the evolution operator. It is found that the frequency domain method needs information of the scattering GF over a much narrower frequency range. In addition, reversible dynamics is observed. These findings are instructive in the fields of coherent light-matter interactions.
\end{abstract}
\volumeyear{year}
\volumenumber{number}
\issuenumber{number}
\eid{identifier}
\date[Date text]{date}
\received[Received text]{date}

\revised[Revised text]{date}

\accepted[Accepted text]{date}

\published[Published text]{date}

\startpage{101}
\endpage{102}
\maketitle

\section{ Introduction}

Recently, considerable attention is devoted to the fields of light-matter
interaction for both fundamental and applicative purposes \cite{cohen1989photons,agarwal1974quantum,peter1994quantum,Buhmann2012Dispersion,Buhmann2012Dispersion2,oulton2009plasmon,Chikkaraddy2016Single,Muskens2007Strong,Kinkhabwala2009Large,Curto930,PhysRevLett.100.203002,doi:10.1021/nl902095q,PhysRevLett.102.146807,Xue2003Decay,Birnbaum2005Photon,PhysRevLett.84.1419,Khitrova2006Vacuum,Yves2007Strong,Noda2007Spontaneous,Altug2006Ultrafast,PhysRevLett.106.196405,PhysRevLett.118.073604,PhysRevB.98.045435,PhysRevLett.119.233901,Zhang:18,doi:10.1021/acs.chemrev.7b00647,doi:10.1021/acs.nanolett.7b01344,kneipp1997single}. Many novel phenomena have been predicted and related devices have been developed, for
example, enhanced and inhibited spontaneous emission, photon blockade,
reversible spontaneous emission, unidirectional emission, trapping atoms by
vacuum forces, enhanced Raman scattering, LEDs, one-atom maser, low-threshold
lasers, etc. Among all, the coherent interaction between a single QE and
its electromagnetic environment is of paramount importance, since it
constitutes one of the most fundamental aspects in QED and is also a practical
need in the fields of quantum information procession as well as
single-molecule sensing. Coherent decay dynamics of an excited QE and the energy level shift are two elementary terms in this field.

Plasmonic nanostructure with an ultrasmall optical mode volume is one of the
most promising platforms for the above two problems, since strong
light-matter interactions at the single-exciton level has been achieved \cite{PhysRevLett.118.237401}. In addition, precise spatial control of the QE position with respect to the plasmonic nanocavity has been realized \cite{Zhang2017Sub}. Further, high-quality metallic nanoparticles with tunable size and controllable shapes can be produced and rationally assembled into well-defined architectures \cite{Tan2011Building,Sheneaap8978}.

Theoretically, the medium-assisted quantized electromagnetic field can be
expressed as the fundamental bosonic vector fields via  photon GF \cite{PhysRevA.68.043816}.
Accordingly, all the information needed to investigate the energy level shift and decay dynamics is contained in the  photon GF. For
example, the energy level shift is expressed by a principal value integration of the electromagnetic photon GF over the whole frequency range \cite{Buhmann2012Dispersion,Buhmann2012Dispersion2}. For nanostructure with high symmetry, such as sphere, the  photon GF can be obtained by semi-analytical methods \cite{tai1994dyadic}. But for nanostructure with arbitrary shape, it is not an easy task. We usually have to resort to numerical methods \cite{doi:10.1002/lpor.201500122,Zhao:18,yunjinwulixuebao,2019arXiv190110078T,PhysRevB.81.125431,Bai:13,PhysRevA.85.053827,VanVlack:12}, such as finite difference time domain method or finite element method, which are time consuming or large memory space requirement.
Since the direct principal value integration method needs information for the photon GF over a wide frequency range,
one treatment is to transform the principle value into an ordinary integration over the imaginary axis, where the Kramers-Kronig (KK) relations of the GF is exploited \cite{PhysRevB.84.075419, Buhmann2012Dispersion2}. In this method, the photon GF as well as the response function for the material with imaginary frequency are needed and hard to be obtained \cite{RevModPhys.88.045003,PhysRevA.76.032106,doi:10.1063/1.4789814}. Differently, we propose a new method by utilizing the subtractive
KK relations without transforming into the imaginary frequency and without worrying about the principal value integral. We will show that this will greatly simplifies the calculation and there is no need of the knowledge for the photon GF over a wide imaginary frequency range.

For the decay dynamics, specially in case of strong coupling regime, coherent decay dynamics exhibit significant reversibility and non-Markovian methods should be taken into account \cite{Rivas_2014}. Usually, this can be numerically resolved by solving a quantum master equation \cite{RevModPhys.88.021002,PhysRevB.92.205420} or the Schr$\ddot{o}$dinger equation \cite{PhysRevA.68.043816,PhysRevB.89.041402,PhysRevA.93.022320} which leads to  the well-known Volterra integral equations of the second kind. Besides the above time domain methods, frequency domain methods based on retarded and advanced Greens function expression for the evolution operator can be used \cite{Xue2003Decay,cohen1989photons}. As explained in Ref. \cite{PhysRevA.85.053827,PhysRevLett.118.237401,PhysRevB.95.161408,PhysRevB.98.045435,Gaveau_1995,Lambropoulos_2000,PhysRevA.99.010102,cohen1989photons}, thit method supplies more obvious physical significance underlying some phenomena and provides distinctive criterion to discriminate between strong coupling and weak coupling. In addition, compared to the time domain method, there is no need of time convolution.

In this work, we first present the theory and derive a general method for calculating the energy level shift. We will show that the energy level shift can be expressed by the sum of the real part of the photon GF and an integral part with the integrand a well-behaved function of the imaginary part of the photon GF. For the following two sections, we apply our method to a particular example where a QE is located around a gold nanosphere. For the energy level shift, we numerically compare our method with the direct Hilbert transformation method and the imaginary frequency integration method \cite{Buhmann2012Dispersion2,PhysRevB.84.075419}. The good performance of our method  will be demonstrated. Then, we show the characteristics of the frequency domain method based on GF and the time domain method of the Volterra integral equation form for calculating dynamics. Section V is devoted to the study of the energy level shift and decay dynamics for QE located in a plasmonic nanocavity. We will demonstrate that frequency domain method for dynamics is more efficient than the time domain method. Finally, a summary is given in Sec. VI.
\section{Model and Method}

Let us consider a two-level QE coupled to a common
electromagnetic reservoir. By using the dipole and rotating-wave
approximations, the total Hamiltonian for the system is \cite{PhysRevA.68.043816}

\begin{align*}
H  &  =H_{0}+H_{I},\\
H_{0}  &  =\int d\mathbf{r}\int_{0}^{+\infty}d\omega\hbar\omega\text{ }%
\hat{\mathbf{f}}^{\dagger}\left(  \mathbf{r},\omega\right)  \cdot
\hat{\mathbf{f}}(\mathbf{r},\omega)+\hbar\omega_{0}|e_{0}\rangle\langle
e_{0}|,\\
H_{I}  &  =-\int_{0}^{+\infty}d\omega\lbrack|e\rangle\langle g|\mathbf{d}^{\ast}
\cdot\hat{\mathbf{E}}\left(  \mathbf{r}_{0},\omega\right)  +\mathbf{H.c}.].
\end{align*}
Here, $\hat{\mathbf{f}}\left(  \mathbf{r},\omega\right)  $ and $\hat
{\mathbf{f}}^{\dagger}\left(  \mathbf{r},\omega\right)  $ are the bosonic
vector field annihilation and creation operators for the elementary excitation
of the electromagnetic reservoir, respectively. $\mathbf{d}=\langle
g|\hat{\mathbf{d}}|e\rangle=d\hat{\mathbf{n}}$ is the element of the
transition dipole moment, with the unit vector $\hat{\mathbf{n}}$ and its strength $d$. The electric field vector operator $\hat{\mathbf{E}}\left(
\mathbf{r},\omega\right)  $ is given by%

\[
\hat{\mathbf{E}}(\mathbf{r},\omega)=i\sqrt{\frac{\hbar}{\pi\varepsilon_{0}}%
}\int d\mathbf{s}\sqrt{\varepsilon_{I}(\mathbf{s},\omega)}\mathbf{G}%
(\mathbf{r},\mathbf{s},\omega)\cdot\hat{\mathbf{f}}(\mathbf{s},\omega)
\]
where $\mathbf{G}(\mathbf{r},\mathbf{s},\omega)$ is the photon GF defined as [$\nabla\times\nabla\times-\varepsilon(\mathbf{r}%
,\omega)\omega^{2}/c^{2}]\mathbf{G}(\mathbf{r},\mathbf{s};\omega)=\omega
^{2}/c^{2}\mathbf{I}\delta(\mathbf{r}-\mathbf{s})$. Here $\varepsilon
(\mathbf{r},\omega)=\varepsilon_{R}(\mathbf{r},\omega)+i\varepsilon
_{I}(\mathbf{r},\omega)$ is the spatially and frequency-dependent complex
relative dielectric function and $\varepsilon_{I}(\mathbf{r},\omega)$ is its
imaginary part. $\mathbf{I}$ is the unit dyad and $c$ refers to the speed of
light in the vacuum.

We assume initially the field is in the vacuum state and the QE is excited. In this case, the states of interest are $|I\rangle=|e\rangle\otimes|0\rangle$ and $|F_{r,\omega}\rangle=|g\rangle\otimes|1_{r,\omega}\rangle$ with $|e\rangle$ ($|g\rangle$) the excited (ground) state of the QE and $|1_{r,\omega}\rangle\equiv\hat{\mathbf{f}}_{j}^{\dagger}\left(
\mathbf{r},\omega\right)  |0\rangle$. $|0\rangle$ is the zero photon state. The time evaluation for this initial state is given by
\begin{align*}
|\Psi(t)\rangle &  \equiv U(t)|I\rangle=c_{1}(t)e^{-i\omega_{0}t}|I\rangle\\
&  +\int dr\int_{0}^{+\infty}d\omega
C(r,\omega,t)e^{-i\omega t}|F_{r,\omega}\rangle.
\end{align*}

The time-dependent Schr\"{o}dinger equation leads to the following equation of
motion for the probability amplitudes:%
\begin{equation}
\overset{\cdot}{c_{1}}(t)=-\int_{0}^{t}K(t-\tau)c_{1}(\tau)d\tau,
\label{Volut}%
\end{equation}
where the kernel function is $K(t-\tau)=\int_{0}^{+\infty}d\omega
J(\omega)e^{i(\omega_{0}-\omega)(t-\tau)}$ with the spectral density defined
$J(\omega)=\operatorname{Im}g_{rr}(\omega)$. Here, the coupling strength
$g_{rr}(\omega)$ is defined by%

\begin{equation}
g_{rr}(\omega)\equiv\frac{\mathbf{d}^{\ast}\cdot \mathbf{G}(\mathbf{\mathbf{r}_{0}}%
,\mathbf{r}_{0},\omega)\cdot\mathbf{d}}{\hbar\pi\varepsilon_{0}}.
\label{couplingstrength}%
\end{equation}

One method for dynamics is to take the time integral on both sides of Eq. (\ref{Volut}). This leads to the well-known Volterra integral equations of the second kind. Explicitly, the probability amplitude for the QE in the excited states is%

\begin{equation}
c_{1}(t)=c_{1}(0)-\int_{0}^{t}B(t-t^{^{\prime}})c_{1}(t^{^{\prime}%
})dt^{^{\prime}} \label{ExplicitVolut}%
\end{equation}
with $B(t-t^{^{\prime}})=\int_{0}^{+\infty}d\omega J(\omega)\int
_{0}^{t-t^{^{\prime}}}e^{-i(\omega-\omega_{0})u}du$, where time-convolution is needed. Note that the kernel $B(t-t^{^{\prime}})$ is an integral over the whole positive frequency spectrum.

Another method for dynamics is by the resolvent operator technique \cite{Lambropoulos_2000,Xue2003Decay,cohen1989photons}. From the GF expression of evolution operator, one can show that $U(t)=\int_{-\infty}^{+\infty}d\omega[G^-(\omega)-G^+(\omega)]exp(-i\omega t)/(2\pi i)$ where $G^\pm(\omega)=\lim_{\eta \rightarrow 0^{+}}G(E\pm i\eta )$ with the resolvent operator $G(z)=(z-H/\hbar)^{-1}$. The probability amplitude $c_{1}(t)$ can be expressed the matrix element of the evolution operator $c_{1}(t)=\langle I|U(t) |I\rangle$. From the operator identity $(z-H_0/\hbar)G(z)=H_I G(z)/\hbar$, we obtain $G_{ii}(\omega)\equiv\langle I|U(t) |I\rangle=(\omega-\omega_0-R_{ii}(\omega))^{-1}$ in which the matrix element of the level-shift operator $R_{ii}(\omega)$ reads
\begin{align*}
R_{ii}(z)=\frac{1}{\pi\varepsilon_{0}}[\int_{0}^{\infty}d\omega\frac
{\mathbf{d}^{\ast}\cdot\operatorname{Im}\mathbf{G}(\mathbf{r}%
_{A},\mathbf{r}_{A},\omega)\cdot\mathbf{d}}{z-\omega}
\end{align*}

Clearly, by using the relation $1/(z-\omega-i\eta)=\wp(1/(z-\omega))+i\pi\delta(z-\omega)$, this can be written

\begin{equation}
R_{ii}^{\pm}(z)=\lim_{\eta \rightarrow 0^{+}}R_{ii}(z\pm i\eta)=\Delta(z)\mp i\frac{\Gamma(z)}{2},
\end{equation}
with
\begin{equation}
\Gamma(z) =2\frac{\mathbf{d}^{\ast}\cdot \operatorname{Im} \mathbf{G}(\mathbf{\mathbf{r}_{0}}%
,\mathbf{r}_{0},z)\cdot\mathbf{d}}{\hbar\varepsilon_{0}}=2\pi\operatorname{Im}g_{rr}(z)\theta(z).
\label{gama}%
\end{equation}
Here, $\theta(z)$ is the step function. $\Delta(z)$ is the Hilbert transform of $\Gamma(z)$, which is
\begin{equation}
\Delta(z)=\frac{1}{2\pi}\wp
\int_{0}^{+\infty}ds\frac{\Gamma(s)}{z-s}.
\label{Hilberttransform}%
\end{equation}

Thus, the probability amplitude reads
\begin{equation}
c_{1}(t) =\lim_{\eta\rightarrow0_{+}}[\int_{-\infty}^{+\infty}%
S(\omega)e^{-i(\omega-\omega_{0})t}d\omega,
\label{Fourier-Laplacedynamics}%
\end{equation}
with evolution spectrum
\begin{equation}
S(\omega)=\frac{1}{\pi}\lim_{\eta\rightarrow0_{+}}\frac{\Gamma(\omega)/2+\eta}{[\omega-\omega
_{0}-\Delta(\omega)]^{2}+(\Gamma(\omega)/2+\eta)^{2}]}.
\label{spectrum}%
\end{equation}

Although the dynamics for our system can be obtained by the above two methods (Eq. (\ref{ExplicitVolut}) and Eq.
(\ref{Fourier-Laplacedynamics})), one has to evaluate the
kernel $B(t-t^{^{\prime}})$ or $\Delta(\omega)$. For both, an integral over the whole frequency range should be performed. However, the Hamiltonian is nonrelativistic
and can not be applied in the case of relativistic high frequency. In addition, both $\Delta(\omega)$ and the kernel $B(t-t^{^{\prime}})$ are divergent. In the
homogeneous case such as vacuum, the divergence can be overcome by the procedure of mass
renormalization. In artificial nanostructure, there is no response
for the medium in the high-frequency range. Thus, one treatment is to separate the homogeneous and scattering contributions, where
the photon GF is decomposed into its bulk and scattered part
$\mathbf{G}(\mathbf{\mathbf{r}_{0}},\mathbf{r}_{0},\omega)=\mathbf{G}%
_{0}(\mathbf{\mathbf{r}_{0}},\mathbf{r}_{0},\omega)+\mathbf{G}_{s}%
(\mathbf{\mathbf{r}_{0}},\mathbf{r}_{0},\omega)$. The scattering photon GF
can be used to take place of the total photon GF where the
homogeneous-medium contribution is attributed to the definition of the
transition frequency of the QE. In this work, we will numerically
confirm that this nature renormalization procedure can overcome the divergence.

In the weak coupling limit where both $\Gamma(\omega)$ and $\Delta(\omega)$
are small comparied with $\omega_{0}$ and vary slowly with frequency $\omega$
around $\omega_{0}$, $\Gamma(\omega)$ and $\Delta(\omega)$ in Eq.
(\ref{spectrum}) can be safely replaced by $\Gamma(\omega)=\Gamma$ and $\Delta(\omega)=\Delta$. Then, the spectrum
$S(\omega)=\frac{1}{\pi}\frac{\Gamma/2+\eta}{[\omega-\omega_{0}-\Delta
]^{2}+(\Gamma/2+\eta)^{2}]}$ is of a Lorentzian form and the result for Eq.
(\ref{Fourier-Laplacedynamics}) is $c_{1}(t)=\exp[-\left(  i\Delta
+\Gamma/2\right)  t]$. This clearly demonstrates that $\Gamma$ and $\Delta$
are the spontaneous emission rate and the energy level shift, respectively. In
the following of this paper, $\Gamma(\omega)$ and $\Delta(\omega)$ are termed
spontaneous emission rate and energy level shift, even though both of them are
maybe highly peaked and vary rapidly with frequency.

The spontaneous emission rate $\Gamma(\omega)$ (Eq. (\ref{gama})) can be obtained once we known the photon GF (see Eq. (\ref{couplingstrength})). But for the
energy level shift $\Delta(\omega)$ (Eq. (\ref{Hilberttransform})), there needs a principle integration which is difficult from a numerical view. Alternatively, there is one method based on the contour-integral techniques, where real
frequency integral is transformed into ones along the positive imaginary axis
plus contributions from the poles \cite{Buhmann2012Dispersion,Buhmann2012Dispersion2,PhysRevB.84.075419}. By utilizing the KK relation of
the photon GF, one has \cite{PhysRevB.84.075419,PhysRevA.66.063810}
\begin{align}
\Delta(\omega)  &  =[\frac{-1}{\hbar\varepsilon_{0}}\mathbf{d\cdot
}\operatorname{Re}\mathbf{G}(\mathbf{\mathbf{r}_{0}},\mathbf{r}_{0}%
,\omega)+\nonumber\\
&  \frac{\omega}{\hbar\pi\varepsilon_{0}}\mathbf{d\cdot}\int_{0}^{+\infty}%
d\xi\frac{\mathbf{G}(\mathbf{\mathbf{r}_{0}},\mathbf{r}_{0},i\xi)}{\omega
^{2}+\xi^{2}}]\cdot\mathbf{d}\nonumber\\
&  =-\pi\operatorname{Re}g_{rr}(\omega)+\omega\int_{0}^{+\infty}d\xi
\frac{g_{rr}(i\xi)}{\omega^{2}+\xi^{2}} \label{imaginarefredelta}%
\end{align}

Besides, we propose a third method by using the subtractive
KK relation. By using the relations $-\pi\operatorname{Re}\mathbf{G}(\mathbf{\mathbf{r}_{0}},\mathbf{r}_{0},s)=\wp\int_{-\infty}^{+\infty}ds\operatorname{Im}\mathbf{G}(\mathbf{\mathbf{r}_{0}},\mathbf{r}_{0},s)/(\omega-s)$ and $\operatorname{Im}\mathbf{G}(\mathbf{\mathbf{r}_{0}},\mathbf{r}_{0},-s)=-\operatorname{Im}\mathbf{G}(\mathbf{\mathbf{r}_{0}},\mathbf{r}_{0},s)$ , the energy level shift for $\omega\geq0$ can be written \cite{PhysRevB.84.075419}%
\begin{align}
\Delta(\omega)  &  =\wp\int_{0}^{+\infty}ds\frac{\operatorname{Im}g_{rr}%
(s)}{\omega-s} \nonumber\\
&  =-\pi\operatorname{Re}g_{rr}(\omega)+\wp\int_{0}^{+\infty}ds\frac
{\operatorname{Im}g_{rr}(s)}{\omega+s}. \label{deltaomega}
\end{align}

For $\omega=0$ , the above equation becomes
\begin{equation}
\Delta(0)  =-\pi\operatorname{Re}g_{rr}(0)+\wp\int_{0}^{+\infty}ds\frac
{\operatorname{Im}g_{rr}(s)}{s}. \label{delta0}
\end{equation}

Subtracting Eq. (\ref{delta0}) from Eq. (\ref{deltaomega}), we have
\begin{equation}
\Delta(\omega)=-\pi\operatorname{Re}g_{rr}(\omega)+\frac{\pi}{2}%
\operatorname{Re}g_{rr}(0)-\omega\int_{0}^{+\infty}ds\frac{\operatorname{Im}%
g_{rr}(s)}{(\omega+s)s}. \label{SubtractiveKK}%
\end{equation}
Here, we have used the relation $\Delta(0)=-0.5\pi\operatorname{Re}g_{rr}(0)$, since the second term on the right hand side in Eq. (\ref{delta0}) is $-\Delta(0)$ ( see Eq. (\ref{Hilberttransform}) ).

Equation (\ref{SubtractiveKK}) is the central result of our method for
evaluating the energy level shift of a QE around arbitrary
nanostructure. In this form, there is no need to worry about the principal
value. Furthermore, for frequency away from the practical resonance where material becomes transparent, the scattering
is weak. Thus, $\operatorname{Im}g_{rr}(s)$ in the integrand is
small, which is very useful in calculating the energy level shift by numerical means.

Before proceeding further, let us give some discussions about the above three
methods to obtain the energy level shift $\Delta(\omega)$. Hereafter, Eq. (\ref{imaginarefredelta}),  Eq. (\ref{Hilberttransform}) and Eq. (\ref{SubtractiveKK}) refer to the imaginary frequency method, the direct Hilbert method and the subtractive KK method respectively. For the imaginary frequency method, real frequency integral is transformed
into the positive imaginary axis, where the photon GF $\mathbf{G}%
( \mathbf{\mathbf{r}_{0}},\mathbf{r}_{0},i\xi )$ decays exponentially with $\xi$
and there is no need to worry about the principle value. However, there is
extra burden to compute the Green tensor for imaginary frequency
$\mathbf{G}(\mathbf{\mathbf{r}_{0}},\mathbf{r}_{0},i\xi)$, compared to the
other two methods ( Eq. (\ref{Hilberttransform}) and Eq. (\ref{SubtractiveKK}%
) ). For nanostructures with high symmetry, $\mathbf{G}(\mathbf{\mathbf{r}_{0}%
},\mathbf{r}_{0},i\xi)$ can be obtained semi-analytically. But for arbitrary
nanostructure which is the usual case, it is a difficult task \cite{RevModPhys.88.045003,PhysRevA.76.032106,doi:10.1063/1.4789814}. Comparing the
two integral part in Eq. (\ref{Hilberttransform}) and Eq.
(\ref{SubtractiveKK}), we find that the integrand in our subtractive KK method ( Eq. (\ref{SubtractiveKK}) )
decays faster than that in the direct Hilbert method ( Eq. (\ref{Hilberttransform}) ) with high frequency
$s$. This can be clearly seen by comparing the integrands. Thus, our subtractive KK method should converge more rapidly than the direct Hilbert method. Further, there is no need to
worry about the principle value as in the imaginary frequency method ( Eq. (\ref{imaginarefredelta}) ). In the next section, we will demonstrate that our subtractive KK method ( Eq. (\ref{SubtractiveKK})) is powerful and can be used to obtain the exact energy level shift efficiently.

\begin{figure}[ptbh]
\centering
\includegraphics[width=6cm]{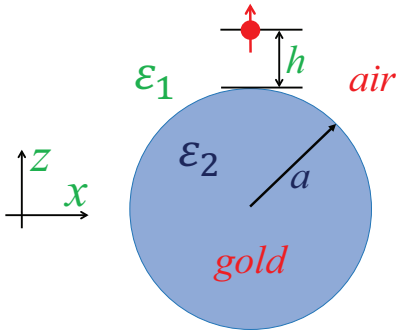}
\caption{Schematic diagrams. A QE is located around a gold nanosphere with radius $a=20nm$. For simplicity, the transition dipole moment for the QE is thought to be polarized along the radial direction. $\varepsilon_1$ and $\varepsilon_2$ are the permittivities for air and gold, respectively.  The distance between the emitter and the surface of metal is $h$}
\label{fig1}%
\end{figure}

As a demonstration, we will apply the methods introduced above to a particular
example where a two-level QE is located above a metal nanosphere
(see Fig. \ref{fig1}). The nanosphere with radius $a$ is located at the origin. A QE at a distance $h$ from the surface of the sphere lies on the x-axis
of the coordinate system. The metal is chosen to be Gold and characterized by
a complex Drude dielectric function \cite{PhysRevB.92.205420} $\varepsilon(\omega)=1-\omega_{p}%
^{2}/\omega(\omega+i\gamma_{p})$ with $\omega_{p}=1.26\times10^{16}$ $rad/s$
and $\gamma_{p}=1.41\times10^{14}$ $rad/s$. The background is vacuum with
$\varepsilon_{B}=1$. For simplicity, the dipole is polarized along the radial
direction of the sphere $\mathbf{d=}d\mathbf{r}$ and its strength is set to $d=24D$. In this case, the photo GF is obtained semi-analytically \cite{339756,tai1994dyadic}

In the following sections, we first demonstrate the validity and advantages of
our method comparing with the other two methods in obtaing the energy level
shift. Then, the properties of the the two methods (Eq. (\ref{ExplicitVolut})
and Eq. (\ref{Fourier-Laplacedynamics})) for evaluating dynamics are
numerically investigated with different truncation conditions. In both cases,
we choose the nanosphere system shown in Fig. 1 as an example, which is widely
investigated and can be treated analytically. We will show that dynamics by
the Fourier-Laplace transformation method with level shift obtained by our
subtractive Kramers-Kronig relation can be numerically evaluated with the
lease computation resource. At the end, we will apply the method introduced
above to investigate the dynamics of a quantum dot in plasmonic nanocavity. Non-Weisskopf-Wigner decay phenomenon will be shown.

\section{NUMERICAL COMPARISION OF THE ABOVE THEREE METHODS FOR ENERGY LEVEL SHIFT}

\begin{figure}[ptbh]
\centering
\includegraphics[width=4cm]{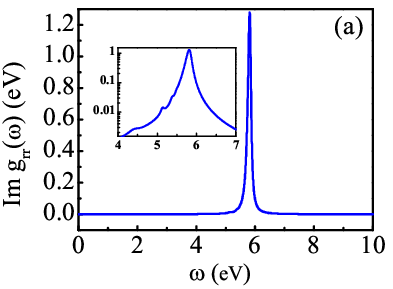}
\includegraphics[width=4cm]{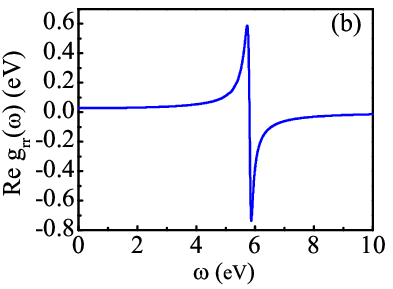}
\includegraphics[width=4cm]{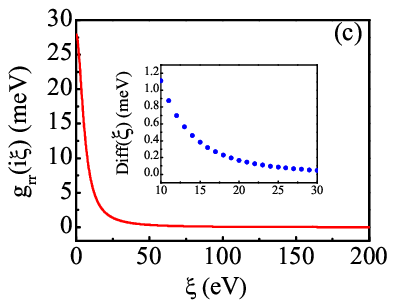}
\includegraphics[width=4cm]{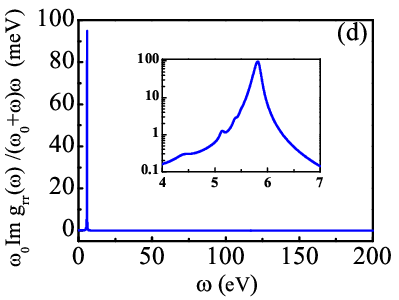}
\caption{Property of the coupling strength $g_{rr}(\omega)\equiv\mathbf{d\cdot G}(\mathbf{\mathbf{r}_{0}},\mathbf{r}_{0},\omega)\cdot\mathbf{d/}\hbar\pi\varepsilon_{0}$ as a function of frequency. (a) The imaginary part $\operatorname{Im}g_{rr}(\omega)$. (b) The real part $\operatorname{Re}g_{rr}(\omega)$. (c) $g_{rr}(\omega)$ with imaginary frequency $\omega=i\xi$. The inset in (c) is $Diff(\xi)=\omega\int_{\xi}^{200eV}ds
g_{rr}(is)/(\omega^{2}+s^{2})$ which is the contribution from $\xi$ to $200eV$. It is in the order of $meV$. (d) The integrand in Eq. (\ref{SubtractiveKK}) $\omega_0\operatorname{Im}g_{rr}(s)/(\omega_0+\omega)\omega$, which shows that integration in our method is well-behaved. Insets in (a) and (d): Zoom of panels showing results in the frequency range (4eV, 7eV) where plasmonic takes great effect. Here, $a=20nm$ and $h=1nm$ for $d=24D$.}
\label{fig2}
\end{figure}

In this section, the above three methods ( Eq.
( \ref{Hilberttransform} ), Eq. ( \ref{imaginarefredelta}) and Eq. ( \ref{SubtractiveKK} ) ) for obtaining the energy level shift are numerically investigated. For nanosphere, the scattering photon GF can be obtained semi-analytically \cite{Zhao:18}. The properties of the coupling strength defined by $g_{rr}(\omega)\equiv\mathbf{d\cdot G}(\mathbf{\mathbf{r}_{0}},\mathbf{r}_{0},\omega)\cdot\mathbf{d/}\hbar\pi\varepsilon_{0}$ ( Eq.(\ref{couplingstrength}) ) are shown in Fig. \ref{fig2} for $a=20nm$ and $h=1nm$. From the inset in Fig. \ref{fig2}(a), we clearly see that there are some peaks for
$\operatorname{Im}g_{rr}(\omega)$ in the frequency range between $4eV$ and
$6eV$, which stem from the localized surface plasmon (LSP) resonance. Over
this frequency range, LSP takes great effect and both the real part and the
imaginary part of $g_{rr}(\omega)$ may be large (see Fig. \ref{fig2}(a) and \ref{fig2}(b)). But
for frequency away, both $\operatorname{Re}g_{rr}(\omega)$
(real part) and $\operatorname{Im}g_{rr}(\omega)$ (imaginary part) are small.
This leads to the usual assumption that higher frequency parts contribute
little to the dynamics or energy level shift.

To evaluate the energy level shift, the first method (Eq.
(\ref{imaginarefredelta})) needs information about the coupling strength
$g_{rr}(i\xi)$ with imaginary frequency $i\xi$. Although it is not an easy
task to obtain the precise photon GF with imaginary frequency in arbitrary
nanostructure, its exact
value can be obtained by replacing the frequency $\omega$ in the semi-analytical expression of $\mathbf{G}(\mathbf{\mathbf{r}_{0}%
},\mathbf{r}_{0},\omega)$ with $i\xi$. Figure \ref{fig2}(c) shows that
$g_{rr}(i\xi)$ is real and decays with $\xi$, which agree well with the
description of Ref. \cite{PhysRevA.76.032106}. It is
well-behaved and the second part in Eq. (\ref{imaginarefredelta}) can be
obtained precisely without difficulty for this particular example. Results from this method are used as a
reference for the other two methods (see Eq.
(\ref{Hilberttransform}) and Eq. (\ref{SubtractiveKK})). Note that $g_{rr}(i\xi)$ spread over a wide frequency range. The inset shows the integral part in Eq. (\ref{imaginarefredelta}) from $\xi$ to $200eV$ $Diff(\xi)=\omega\int_{\xi}^{200eV}ds
g_{rr}(is)/(\omega^{2}+s^{2})$ for $\omega=5eV$. It is about $1.1meV$ for $\xi=10eV$.

Figure \ref{fig2}(d) shows the integrand $\omega_0\operatorname{Im}g_{rr}(\omega)/(\omega_0+\omega)\omega$ as a function of frequency $\omega$ in our subtractive KK method ( see Eq.
(\ref{SubtractiveKK})) with $\omega=4.41eV$, which is around the LSP dipole
mode. Also shown, in the inset of Fig. \ref{fig2}(d), is a plot over the frequency range where LSP takes great effect. We find that this integrand is also well-behaved and can be handled easily and precisely. This is true for other transition frequency $\omega$. Note that there is no need about the knowledge of the photon GF on the imaginary axis, which is different from the first method. In addition, the integrand in our method (Eq. (\ref{SubtractiveKK})) is peaked over a narrower frequency range around the resonance frequency of the nano-sphere, which is different from the integrand in Eq. (\ref{imaginarefredelta}) (see Fig. \ref{fig2}(c) where it remains over a much wider frequency range).

\begin{figure}[ptbh]
\centering
\includegraphics[width=4cm]{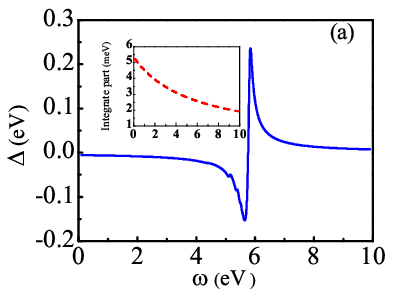}
\includegraphics[width=4cm]{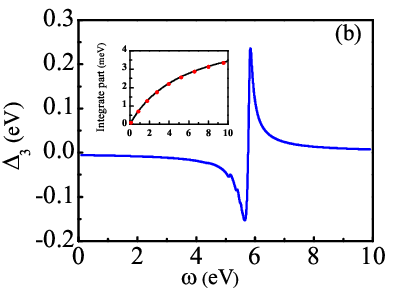}
\includegraphics[width=4cm]{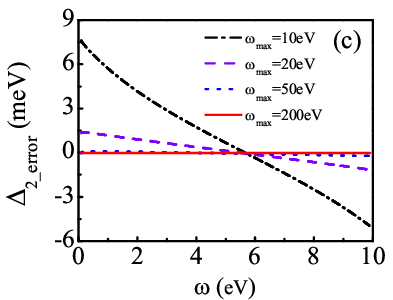}
\includegraphics[width=4cm]{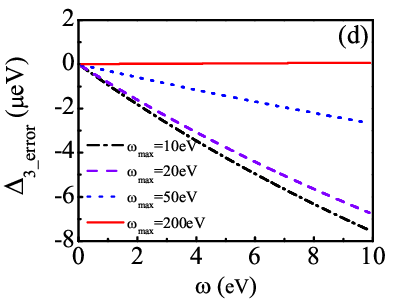}
\caption{Performances of the three different methods for calculating the energy level shift. (a) $\Delta(\omega)$ obtained by the imaginary frequency method. The integral part in Eq. (\ref{imaginarefredelta}) over $(0eV,200eV)$ is shown in the inset. (b) $\Delta(\omega)$ obtained by our subtractive KK method. The inset is for the integral term in Eq. (\ref{SubtractiveKK}) ( The black solid line and red dot are for the integral on the range $(0eV,200eV)$ and $(3eV,8eV)$, respectively. It is found that they agree well.). (c) and (d) are for $\Delta_{2\_error}(\omega)$ and $\Delta_{3\_error}(\omega)$ respectively with $\Delta_{i\_error}(\omega) \equiv \Delta(\omega)-\Delta_i(\omega)$ where $\Delta_2(\omega)$ and $\Delta_3(\omega)$ are obtained from Eq. (\ref{Hilberttransform}) and Eq. (\ref{SubtractiveKK}) respectively. Note that the units in (c) and (d) are $meV$ and $\mu eV$ respectively. Here, $a=20nm, h=2nm$. }
\label{fig3}%
\end{figure}

The energy levle shift obtained by the above three methods are shown in Fig.
\ref{fig3}. From a computational perspective, the upper limit of integration $+\infty$
should be replaced by some cut-off frequency $\omega_{max}$. Figure 3(a)
shows the energy level shift $\Delta(\omega)$ by the first method (Eq.
(\ref{imaginarefredelta})), where $\omega_{\max}=200eV$. The results are
convergent, as the difference between $\omega_{\max}=100eV$ and $\omega_{\max}=200eV$ is less than $10^{-7}eV$. The inset shows the integral term for the imaginary frequency method ( Eq. (\ref{imaginarefredelta})) ). Although, it is in the range of a few $meV$ for parameters considered here, it can not be overlooked where its appearance and importance are discussed in Ref. \cite{PhysRevB.84.075419}.

As a demonstration of our subtractive method, Fig. \ref{fig3}(b) shows the results from Eq. (\ref{SubtractiveKK}) with the integral part ( the last term ) shown in the inset. The black solid line and the red dots are results for integral on the range between $(0eV,200eV)$ and $(3eV,8eV)$, respectively. It is found that they agree well, which means that our method needs knowledge over a narrow frequency range which is different from the imaginary frequency method (see Fig. 2(c) and inset therein). In addition, this term is also in the range of a few $meV$. Different from that shown in the inset in Fig. 3(a), it grows with increasing transition frequency. In the visible range, this integral part is lower than that in the inset of Fig. 3(a). For example, at $\omega=2eV$, it is below $1.5meV$, while it is above $3.5meV$ for the imaginary frequency (see the inset in Fig. 3(a)).

To test the accuracy, result from the imaginary frequency method ( Eq. (\ref{imaginarefredelta})) is thought to be precise and served as a reference. We define the relative errors $\Delta_{i\_error}%
(\omega)=\Delta_{i}(\omega)-$ $\Delta(\omega)$ $(i=2,3)$, where $\Delta
(\omega)$ is the energy level shift obtained by the imaginary frequency method (results in Fig. \ref{fig3}(a) by Eq. (\ref{imaginarefredelta})) and $\Delta_{2}(\omega)$
($\Delta_{3}(\omega)$) represents the results by the direct Hilbert method ( our subtractive KK method )
through Eq. (\ref{Hilberttransform}) ( Eq. (\ref{SubtractiveKK}) ). Numerically,
the upper limit of both integrations should be truncated to some value
$\omega_{\max}$. Figure \ref{fig3}(c) are the results for $\Delta_{2\_error}(\omega)$ with $\omega_{\max}=10eV$ (black dash dot), $20eV$ ( purple dash ), $50eV$ ( blue
dot ) and $200eV$ ( red solid line ). We find that $\left\vert \Delta_{2\_error}(\omega)\right\vert $  in the frequency range is a few $meV$, when the cut-off frequency is less than
$\omega_{\max}=20eV$. A much higher value for $\omega_{\max}$, for example, $\omega_{\max}>20eV$ is needed in order to meet $\left\vert \Delta_{2\_error}(\omega)\right\vert < 1meV$ . But for our subtractive method, $\left\vert \Delta_{3\_error}(\omega)\right\vert $ shown in Fig. \ref{fig3}(d) are less than $0.008 meV$ for all the above four different cut-off frequency $\omega_{\max}$. This clearly confirms that knowledge of the photon GF over a narrow frequency range is enough to get the precise energy level shift by our subtractive KK method ( Eq. (\ref{SubtractiveKK}) ).

It should be stressed that numerical evaluation of the photon GF is not
an easy task for arbitrary nanostructure. Recently, we have proposed a finite
element method to exactly calculate the scattering photon GF. For one
frequency point, two different runs are needed. In addition, for a
single run, the typical computational time is about half an hour on
our\ workstation with processor "Intel(R) Xeon(R) E5-2697 v3" and the memory
usage is about $30GB$, where the simulation domain has been
even reduced to one quarter for symmetry consideration of the nanosphere
system. Since the frequency range by our method is much narrower than by
the direct Hilbert method, we can conclude that our method is much better. In addition, once the dynamics is calculated by the frequency domain method through Eq. (\ref{Fourier-Laplacedynamics}), one has to resolve the photon GF on the imaginary axis. Compared to our subtractive KK method where knowledge of photon GF with real frequency is enough, the imaginary frequency method needs extra simulation.

\begin{figure}[ptbh]
\centering
\includegraphics[width=4cm]{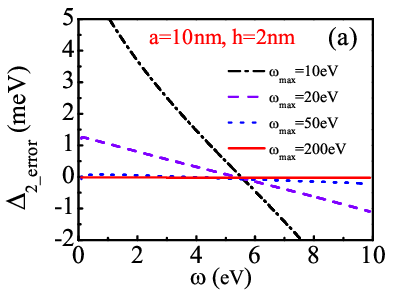}
\includegraphics[width=4cm]{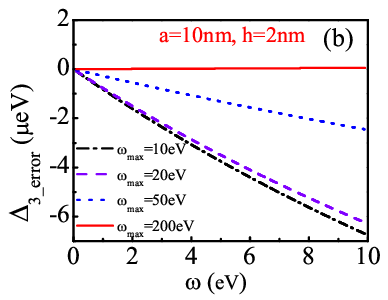}
\includegraphics[width=4cm]{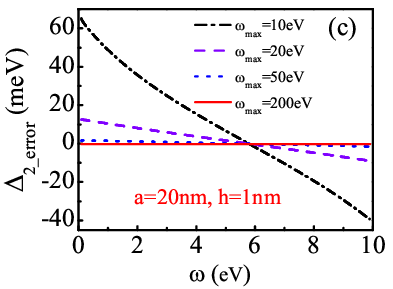}
\includegraphics[width=4cm]{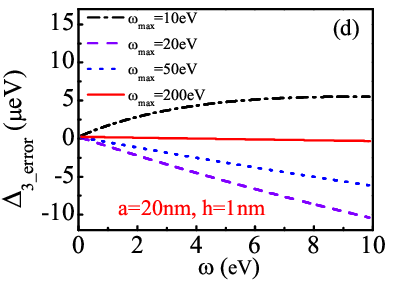}
\caption{Performances of the three different methods for calculating the energy level shift. (a) and (c) are for $\Delta_{2\_error}(\omega)$ by Eq. (\ref{imaginarefredelta}) in units of $meV$. (b) and (d) are for $\Delta_{3\_error}(\omega)$ by Eq. (\ref{SubtractiveKK}). Note that the units here are $\mu eV$. (a) and (b) are for $a=10nm, h=2nm$. (c) and (d) are for $a=20nm, h=1nm$.}
\label{fig4}
\end{figure}

To further demonstrate the good performance of our subtractive KK method, we also investigate the
case for $a=10nm, h=2nm$ and $a=20nm, h=1nm$. Figure \ref{fig4}(a) and \ref{fig4}(b) show similar phenomena as those in Fig. \ref{fig3}(c) and \ref{fig3}(d), where the relative error is a few $\mu eV$ for our method and is a few $meV$ for the direct Hilbert method with $\omega_{max}\leq 20eV$. The relative error for the direct Hilbert method increases with the emitter approaching the surface of nanosphere, which can be clearly seen by comparing Fig. \ref{fig4}(c) with Fig. \ref{fig4}(a) or Fig. \ref{fig3}(c). However, the relative error for our method remains at a extremely low level, for example, less than $10 \mu eV$ for $\omega_{max}\geq 10eV$.

From the above results, we can conclude that Eq. (\ref{SubtractiveKK}) is an
efficient method to obtain the energy level shift of a QE in arbitrary
nanostructure, where the argument of the frequency in photon GF is real.
The integration can be made by common techniques and converges much quickly.

\section{CHARACTERISTICS OF THE TWO METHODS FOR DYNAMICS}
In this section, we numerically demonstrate the characteristics for the above two methods shown in Eq. (\ref{ExplicitVolut}) and Eq. (\ref{Fourier-Laplacedynamics}) used in calculating dynamics. For simplicity, the parameters about the system are the same as those in Fig. \ref{fig4}(c) and \ref{fig4}(d). The transition frequency is set to $\omega_{0}=5eV$. Figure 5(a) are the results by the time domain method through solving the Volterra integral equation ( Eq. (\ref{ExplicitVolut}) ) with the cut-off frequency $\omega_{max}=10eV$ ( blue solid line ) and $\omega_{max}=20eV$ ( red circles ). We see that results are the same for both cases, which means
that a small cut-off frequency, for example, $\omega_{max}=10eV$ is enough to
obtain a convergent result for this particular example.

\begin{figure}[htbp]
\centering
\includegraphics[width=8cm]{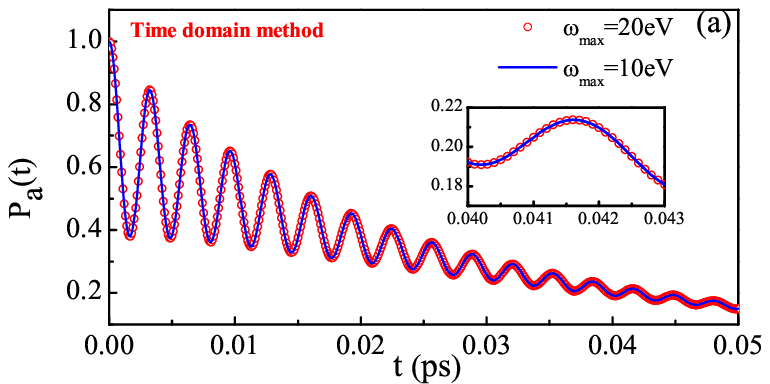}
\includegraphics[width=8cm]{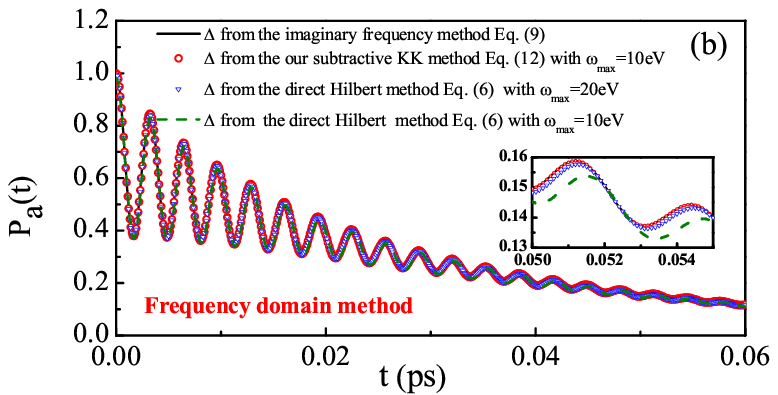}
\includegraphics[width=8cm]{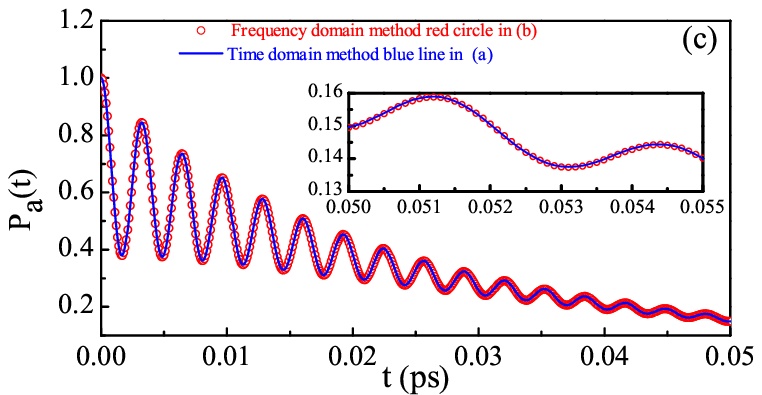}
\caption{Characteristics for the time domain method and for frequency domain method in the decay dynamics of the excited-state population $P_a(t)=|c_1(t)|^2$ . (a) Results by the time domain method by solving Eq. (\ref{ExplicitVolut}). The blue line ( red circle ) is for cut-off frequency $\omega_{max}=10eV$ ( $\omega_{max}=20eV$ ). The inset is for relative longer times. It is found that they agree very well. (b) Results by the frequency domain method through Eq. (\ref{Fourier-Laplacedynamics}) with $\Delta(\omega)$ obtained by imaginary frequency method Eq. (\ref{imaginarefredelta}) ( black solid line ), our subtractive KK method Eq. (\ref{SubtractiveKK}) ( red circle for $\omega_{max}=10eV$ ) and direct Hilbert method Eq. (\ref{Hilberttransform}) ( blue triangles for $\omega_{max}=20eV$ and green dashed line for $\omega_{max}=10eV$ ). The inset is for relative longer times, from which we see that results with $\Delta(\omega)$ obtained from Eq. (\ref{Hilberttransform}) deviate from those with $\Delta(\omega)$ obtained from Eq. (\ref{SubtractiveKK}) and Eq. (\ref{imaginarefredelta}). (c) Comparison of the results by Eq. (\ref{ExplicitVolut}) ( blue solid line in (a) ) and Eq. (\ref{Fourier-Laplacedynamics}) (red circle in (b) ). Here, the transition frequency $\omega_0=5eV$ and the other parameters are the same as those in Fig. 4(c) and 4(d).}
\label{fig5}
\end{figure}

For the frequency domain method based on Green's function expression for the evolution operator ( Eq. (\ref{Fourier-Laplacedynamics}) ), we need the knowledge of the energy level
shift $\Delta(\omega)$. In the above section, we have shown that our method
converges more quickly than the direct Hilbert transform method ( Eq. \ref{Hilberttransform} ). In addition,
there are some minor errors for relative low cut-off frequency. Figure 5(b) shows how these errors
influence the dynamics of the emitter. From the inset, we find that results when $\Delta$ is obtained from the direct Hilbert method ( Eq. \ref{Hilberttransform} ) deviate from those when $\Delta(\omega)$ is obtained by the imaginary frequency domain method ( Eq. (\ref{imaginarefredelta}) ) or our subtractive KK method ( Eq. (\ref{SubtractiveKK}) ). In addition, the larger the cut-off frequency $\omega_{max}$ is, the less the error is.

Thus, the dynamics by the frequency domain method based on the Green's function expression of the evolution operator is sensitive to the
energy level shift $\Delta(\omega)$, which implies that $\Delta(\omega)$
should be precise to obtain a convergent results. In addition, Fig. 5(c) shows that both methods produce the same results as long as $\Delta(\omega)$ is convergent.
This phenomenon remains for different parameters, such as the transition
frequency $\omega_0$, the location of the emitter and the radius of the nanosphere, which are not shown here.

Although the dynamics of an excited dipole can be addressed by both methods in this example, the frequency domain method (Eq. (\ref{Fourier-Laplacedynamics})) provides direct information about the spontaneous emission spectrum, which is helpful to understand the time evolution of the system in various regimes (Detail discussion can be found in Chapter III of Ref. \cite{cohen1989photons}, where the spontaneous emission spectrum changes progressively from a Lorentizan form to a set of two delta functions are demonstrated.). Recently, bound state with decoherence dynamics in photonic crystal \cite{PhysRevLett.64.2418,PhysRevA.50.1764,doi:10.1021/acsphotonics.8b01455,Liu2016Quantum}, cavity arrays \cite{PhysRevA.89.053826} and plasmonic nanostructure have been discussed. To form a bound state, there should be a discrete eigenstate with eigenenergy in the photonic band gap or in the negative axis $\omega_b <0$, where $\omega_b$ is one root for equation $\omega-\omega_{0}-\Delta(\omega)=0$. With this in hand, it is helpful to explain the above phenomena and it is instructive to classify the coupling in various regimes. In addition, we will demonstrate in the next section that a much narrower frequency range for the photon GF is needed by the frequency domain method, when more realistic permittivity beyond the Drude model assumption for material is taken into account.

The main results of this section can be summarized as follows. Decay dynamics can be obtained by the time domain method by solving the Volterra integral equation ( Eq. (\ref{ExplicitVolut}) ) or by the frequency domain method based on the Green's function expression of the evolution operator ( Eq. (\ref{Fourier-Laplacedynamics}) ) with the energy level shift $\Delta(\omega)$ calculated by our subtractive KK method ( Eq. (\ref{SubtractiveKK}) ), in which information over a narrow real frequency range for the photon GF is enough. If the energy level shift $\Delta(\omega)$ is calculated by the imaginary frequency method ( Eq. (\ref{imaginarefredelta})) or by the direct Hilbert method ( Eq. (\ref{Hilberttransform})), knowledge of the photon GF either over a wide imaginary frequency range or a wide real frequency range is demanded to ensure a convergent result.

\section{LEVEL SHIFT AND DYNAMICS OF A QUANTUM EMITTER IN A PLASMONIC
NANO-CAVITY}

\begin{figure}[htbp]
\centering
\includegraphics[width=6cm]{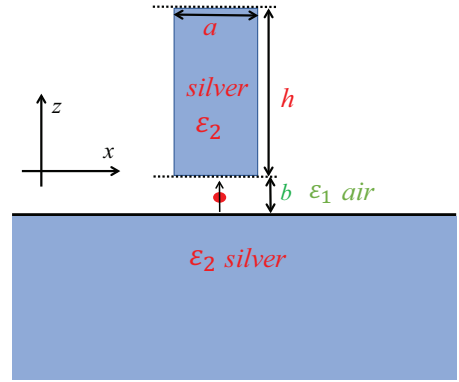}
\caption{Scheme diagram for emitter-nanocavity coupling system in the xoz plane.}
\label{fig6}
\end{figure}

In this section, we apply Eq. (\ref{SubtractiveKK}), Eq.
(\ref{Fourier-Laplacedynamics}) and Eq. (\ref{ExplicitVolut}) to investigate the level shift and dynamics of a QE in a plasmonic nanocavity (see Fig. \ref{fig6}). Here, the nanocavity is composed of a silver nanorod above a silver substrate with a gap distance $b=3 nm$. The diameter and height of the nanorod are $a=10nm$ and $h=30nm$ respectively. For simplicity, a QE with transition dipole moment $d=72D$ polarized along the z-axis is at the center of the gap. Permittivity for the air is $\varepsilon_1=1$. Different from the previous example where Drude model is used for the permittivity of metal over the whole frequency range, permittivity for silver $\varepsilon_2$ is from experiment data \cite{Palik} which is beyond the Drude model.

The coupling strength $g$ is obtained by COMSOL Multiphysics software with the method in Ref. \cite{Zhao:18}, where the scattering GF is expressed by the difference of the electric fields of an oscillating electric point dipole with and without nanostructre. The real part and imaginary part for the coupling strength $g$ are shown in Fig. 7 (a) and 7(b) respectively. Different from the results shown in Fig. 2(a) and Fig. 2(b) where the coupling strength $g$ is near zero for frequency away from the plasmonic resonance, there are some response for relative large frequency (see the inset in Fig. 7(b)). Thus, to evaluate the third part in our subtractive KK method $g(0)$ (Eq. (\ref{SubtractiveKK})) $\Delta_{cor}(\omega)=\omega\int_{0}^{\omega_{max}}ds\operatorname{Im}g_{zz}(s)/(\omega+s)s$ , we should choose a relative large cut-off frequency $\omega_{max}$. Figure 7(c) shows $\Delta_{cor}(\omega)$ with $\omega_{max}=50eV$ (black solid line) and $\omega_{max}=10eV$ (red dashed line). Their difference is shown in the inset. We find that their difference is about $10meV$ for $\omega\approx2eV$. For the second part in our subtractive KK method $g(0)$ (Eq. (\ref{SubtractiveKK})), linearly extrapolating method is used. Results for $Re g(\omega)$  with $\omega$ in the range $[0.125eV,0.2eV]$ is shown in Fig. 7(d). A liner function $Re g_i(\omega)=178.685+2.44152\omega$ is obtained. This result agrees well with that when silver is thought to be perfect conductor.

\begin{figure}[htbp]
\centering
\includegraphics[width=4cm]{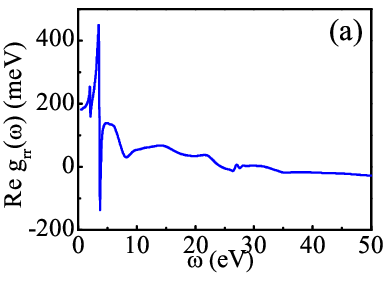}
\includegraphics[width=4cm]{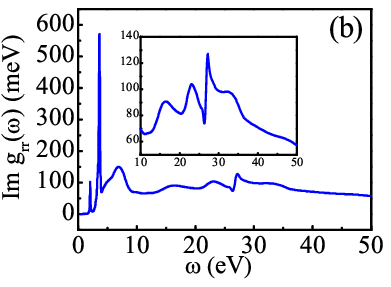}
\includegraphics[width=4cm]{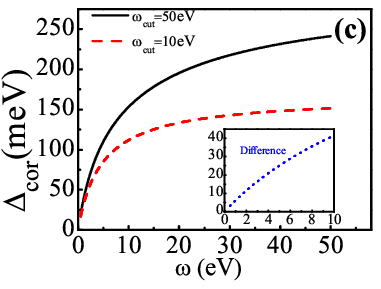}
\includegraphics[width=4cm]{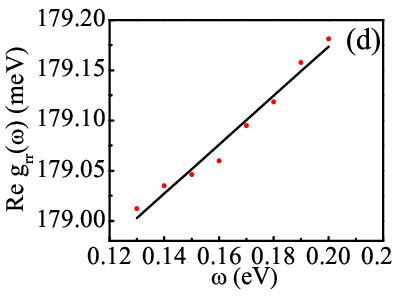}
\includegraphics[width=8cm]{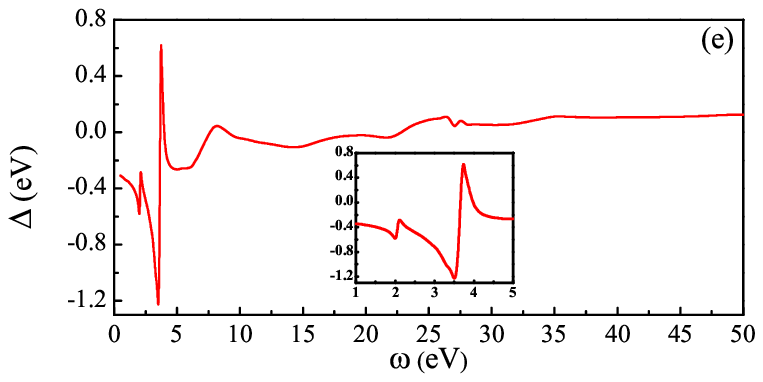}
\caption{Energy level shift $\Delta(\omega)$ and related components at the right hand of Eq. (\ref{SubtractiveKK}). (a) and (b) the real and imaginary part for the coupling strength $g(\omega)$ respectively. (c) the third part in Eq. (\ref{SubtractiveKK}) $\Delta_{cor}(\omega)=\omega\int_{0}^{\omega_{max}}ds\operatorname{Im}g_{zz}(s)/(\omega+s)s$ with $\omega_{max}=50eV$ (black solid) and $\omega_{max}=10eV$ (red dashed). The inset is the difference between $\omega_{max}=50eV$ and $\omega_{max}=10eV$. (d)  $g(\omega)$ with $\omega\rightarrow0$. (e) the energy level shift $\Delta(\omega)$. Insets in (a) and (e): Zoom of panels showing results in the frequency range (1eV, 5eV) where plasmonic takes great effect. }
\label{fig7}
\end{figure}

Figure 8 shows the performances for the time domain method and for frequency domain method. The transition frequency is $\omega_0=2.5eV$. Figure 8(a) are results by the frequency domain method (Eq. (\ref{Fourier-Laplacedynamics})). The black solid line is for the cut-off frequency $\omega_{max}=50eV$ while the red circle is for $\omega_{max}=10eV$. We find that they agree well (see the inset therein). This means that knowledge about the scattering GF over a narrow frequency range is enough. But for the results by the time domain method (Eq. (\ref{ExplicitVolut})) shown in Fig. 8(b), we find that results with $\omega_{max}=20eV$ (red square) differs much from that with $\omega_{max}=50eV$ (black solid). This means that $\omega_{max}=20eV$ is not enough for this method. To compare the results by the above two methods, we show the results by both methods with a relative high cut-off frequency $\omega_{max}=50eV$ in Fig. 8(c). They agree with each other. Thus, we can conclude that a much narrow frequency range for the photon GF is enough to get a convergent results by the frequency domain method than by the time domain method.
\begin{figure}[thbp]
\centering
\includegraphics[width=8cm]{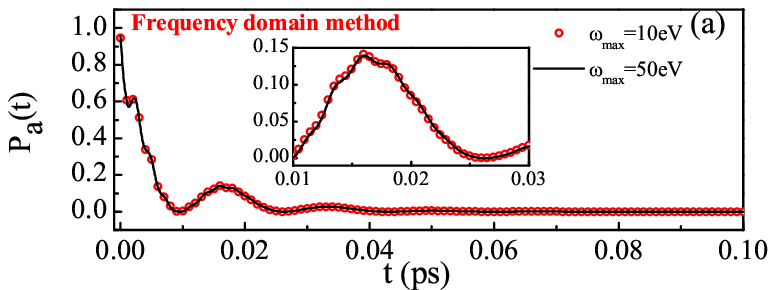}
\includegraphics[width=8cm]{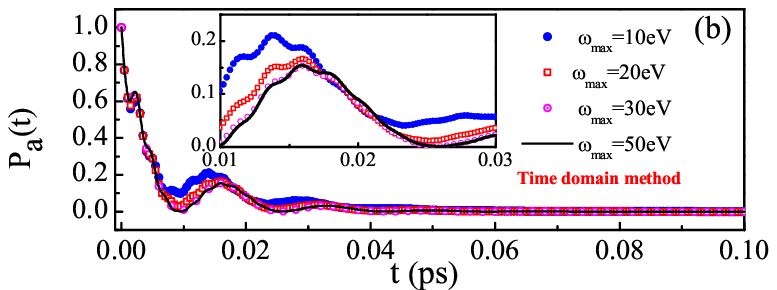}
\includegraphics[width=8cm]{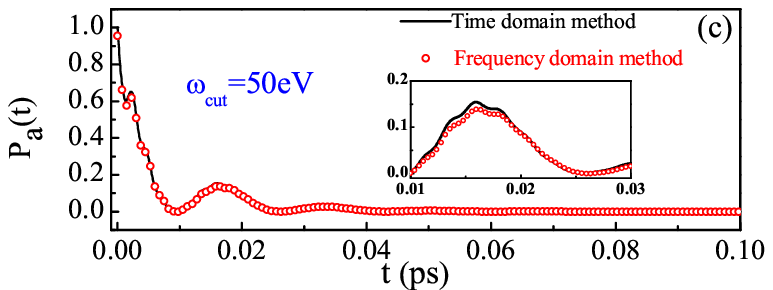}

\caption{Performances for the time domain method and for frequency domain method in the decay dynamics of the excited-state population $P_a(t)=|c_1(t)|^2$ . (a) and (b) are results by the frequency domain method through Eq. (\ref{Fourier-Laplacedynamics}) with $\Delta(\omega)$ obtained by our subtractive KK method (Eq. (\ref{SubtractiveKK})) and by the time domain method through solving Eq. (\ref{ExplicitVolut}) respectively. (c) is for their comparison with $\omega_{max}=50eV$. It is found that the decay dynamics can be obtained by the frequency domain method with a much lower cut-off frequency. Insets: Zoom of panels showing their difference.}
\label{fig8}
\end{figure}

 \section{SUMMARY}
In summary, we have proposed a general numerical method for calculating the energy level shift of a QE in arbitrary nanostructure. By subtracting the expression for the energy level shift $\Delta(0)$ and using the  Kramers-Kronig relations for the scattering photon GF, we have shown that the principal value integral in calculating $\Delta(\omega)$ is transforms into an ordinary integration(see Eq. (\ref{SubtractiveKK})). We have made numerical comparisons with the method of direct Hilbert transformation
over the positive frequency axis (see Eq. (\ref{Hilberttransform})) and the method by transferring the integration to the imaginary frequency axis (see Eq. (\ref{imaginarefredelta})) for emitter located around a gold nanosphere and at the center of a gap plasmonic nonacavity. In the gold nanosphere case, permittivity for metal is supposed to be Drude model, although it can not be extended to a wide frequency range in reality. By the method of integration on the imaginary frequency axis (Eq. (\ref{imaginarefredelta}))), $g(i\xi)$ spread over a wide frequency range and contribution from $\xi>10eV$ is in the order of $meV$. For the method of direct Hilbert method (Eq. (\ref{Hilberttransform})), errors are also in the orders of $meV$ with a cut-off frequency $\omega_{max}=10eV$. But for our method, numerical errors is several $\mu eV$ with a cut-off frequency $\omega_{max}=10eV$. This clearly demonstrate that a much narrower frequency range about the scattering GF is enough for our subtractive KK method (Eq. (\ref{SubtractiveKK})), which is very useful when calculating the GF by numerical means. In addition, we have demonstrated that dynamics by the frequency domain method based on the Greens function expression for the evolution operator sensitively depends on the energy level shift.

For the gap plasmonic nonacavity case, permittivity for the metal is from experiment data which is beyond the Drude model. We have found that the coupling strength $Im g(\omega)$ is relatively strong and response can not be ignored in the high frequency range. We have found that the integral part over the frequency range $[10eV,50eV]$ of our subtractive method (Eq. (\ref{SubtractiveKK})) is about $10meV$ for $\omega$ around $2eV$. Nevertheless, dynamics by the frequency domain method is less affected in the high frequency range. There is no visible difference for the dynamics of the excited-state population between  $\omega_{max}=10eV$ and $\omega_{max}=50eV$. Differently, the time domain method in the form of Volterra integral of the second kind is strongly is affected. We have found that cut-off frequency with $\omega_{max}=20eV$ is not enough to get a convergent results. In addition, we have observed that results gradually approached to those by the frequency domain method with the cut-off frequency $\omega_{max}$ increasing.

\begin{acknowledgments}
 This work was financially supported by the National Natural Science Foundation of China (Grants No. 11464014, 11347215, 11564013, 11402096, 11464013) and Hunan Provincial Innovation Foundation For Postgraduate (Grants No.CX2018B706).
 \end{acknowledgments}


\end{document}